\documentclass[runningheads]{llncs}
\usepackage{graphicx}
\usepackage[nobiblatex]{xurl}
\usepackage{comment}
\usepackage{rotating}
\usepackage{adjustbox}

\usepackage{amsfonts}

\usepackage{amsmath,amssymb,amsfonts}
\usepackage{manyfoot}
\usepackage{booktabs}
\usepackage{graphics}
\usepackage{bigstrut}

\usepackage{url}
\usepackage{multicol}
\usepackage{multirow}
\usepackage{verbatim}
\usepackage{tabularx}
\usepackage{color,soul}
\usepackage{array}
\usepackage{makecell}
\usepackage{hyperref}
\usepackage{subfigure}

\newcommand{\cmt}[1]{}

 \graphicspath{ {./APA/} }

\usepackage[
backend=biber,
style=numeric,
sorting=none
]{biblatex}
\begin{document}
\title{Coronary Artery Disease Classification with Different Lesion Degree Ranges based on Deep Learning}
\titlerunning{CAD Classification with Different Lesion Degree Ranges based on DL}
%
\author{Ariadna Jim\'enez-Partinen\inst{1,3} \and
Karl Thurnhofer-Hemsi\inst{1,3,4} \and
Esteban J. Palomo\inst{1,3} \and
Jorge Rodríguez-Capitán\inst{2,3,4} \and
Ana I. Molina-Ramos\inst{2,3,4}}
\authorrunning{Ariadna Jim\'enez-Partinen et al.}
%
\institute{Department of Computer Languages and Computer Science. University of M\'alaga, Bulevar Louis Pasteur, 35, M\'alaga, Spain, 29071 \and
Cardiology Deparment, Hospital Universitario Virgen de la Victoria, M\'alaga, 29010, Spain\\ \and
Instituto de Investigación Biomédica de Málaga y Plataforma en Nanomedicina-IBIMA Plataforma BIONAND, C/ Severo Ochoa, 35, Málaga TechPark, Campanillas, 29590, M\'alaga, Spain\\ \and
Centro de Investigaci\'on Biom\'edica en Red de Enfermedades Cardiovasculares (CIBERCV), Instituto de Salud Carlos III (ISCIII), Avenida Monforte de Lemos, 3-5. Pabell\'on 11. Planta 0, 28029, Madrid, Spain\\}
\maketitle              
\begin{abstract}
Invasive Coronary Angiography (ICA) images are considered the gold standard for assessing the state of the coronary arteries. Deep learning classification methods are widely used and well-developed in different areas where medical imaging evaluation has an essential impact due to the development of computer-aided diagnosis systems that can support physicians in their clinical procedures. In this paper, a new performance analysis of deep learning methods for binary ICA classification with different lesion degrees is reported. To reach this goal, an annotated dataset of ICA images that contains the ground truth, the location of lesions and seven possible severity degrees ranging between 0\% and 100\% was employed. The ICA images were divided into ``lesion'' or ``non-lesion'' patches. We aim to study how binary classification performance is affected by the different lesion degrees considered in the positive class. Therefore, five known convolutional neural network architectures were trained with different input images where different lesion degree ranges were gradually incorporated until considering the seven lesion degrees. Besides, four types of experiments with and without data augmentation were designed, whose F-measure and Area Under Curve (AUC) were computed. Reported results achieved an F-measure and AUC of 92.7\% and 98.1\%, respectively. However, lesion classification is highly affected by the degree of the lesion intended to classify, with 15\% less accuracy when $<$99\% lesion patches are present.

\end{abstract}

\section{Introduction}
\label{sec:intro}

Invasive Coronary Angiography (ICA) images are one of the methods for anatomical imaging evaluation. Although the use of other non-invasive methods of assessment for Coronary Artery Disease (CAD) is increasing, it remains the gold standard method for evaluating the coronary artery state, confirming CAD, and guiding for interventions through X-ray imaging technology \cite{collet_2020_2021,knuuti_2019_2020,zhou2021review}. During an ICA procedure, a catheter is inserted by a percutaneous incision in the radial or femoral artery to introduce the radiocontrast agent \cite{rigatelli2022modern}. 

The assessment of the stenosis severity is done visually and has a crucial subjective part that depends on the experience of the expert, having a substantial interobserver variability \cite{leape2000effect,zir1976interobserver}. Computer-aided diagnosis could improve the efficiency of diagnosis, supporting clinician decisions. This fact motivates the scientific community to develop and analyze different approaches to solve stenosis classification and detection tasks in ICA images. Nowadays, Convolutional Neural Networks (CNN) use the power of GPU-accelerated algorithms to recognize objects successfully and have been widely used for decision support systems and image classification, more specifically in medical images \cite{cai2020review, zhou2021reviewDL}.

In this context, only a few methods are proposed, being this research field is in an early stage because of the need for available open-access datasets \cite{litjens2019state}. To relieve this problem, Ovalle \textit{et al.} \cite{ovalle2022improving} proposed a Bezier-based Generative Model, which generates synthetic image patches as a data augmentation technique. In order to detect single severe lesions ($\geq$ 70\% of narrowing) in ICA images, a comparison among eight detector architectures considering both detection metrics and real-time data processing was presented by Danilov \textit{et al.} \cite{danilov2021real}, where the architecture based on Faster-RCNN Inception ResNet V2 was the most accurate single-vessel detector.

 Pang \textit{et al.} \cite{pang2021stenosis} designed a two-stage network as an object detector based on ResNet-50 structure that was developed using sequence image information from single projection ICA images. Firstly, a feature map was extracted and candidate boxes were generated and classified into stenosis or non-stenosis in the second stage. A method based on keyframes selection and classification into normal ($<$50\% narrowing) and abnormal ($\geq$ 50\% of narrowing) images using a GoogleNet Inception-V3 as based architecture was proposed by Moon \textit{et al.} \cite{moon2021automatic}. The location of the stenosis was also provided. Zhou \textit{et al.} \cite{zhou2021automated} used a three-stage method for extracting keyframes using ResNet-18 structure, vessel segmentation with U-Net model, and stenosis measurement from segmentation masks to classify Right Coronary Artery (RCA) images according to the lesion degree presented was proposed.

The main contribution of this work is to evaluate how the binary classification performance of ICA images is affected by the different lesion degrees considered in the ``lesion'' class, whose effects have been unreported before in the literature. In addition, a comparison between well-known deep neural network models is analyzed to determine the most effective model for the different lesion degree ranges. This is an exhaustive study to increase understanding of shortcomings, requirements, and potential improvements for deep learning solutions in invasive coronary angiography, approaching solutions for clinical settings, having the potential to alleviate pressure on healthcare services in general and to improve the catheterization laboratory diagnoses, treatment, and logistics in particular, as described below.

Firstly, improving understanding of shortcomings of coronary stenoses can facilitate operators to identify lesions that might have otherwise been unnoticed, which would have a beneficial impact on patient outcome.
Secondly, recording and reporting the results of procedures (such as lesion location, severity and whether stents have been placed) through automated ICA interpretation shortens the duration and increases the efficiency of ICA. This leads to higher amortization of catheterization laboratory.
Thirdly, this study and further work are focus on find and apply models might also be used to guide real-time Percutaneous Coronary Intervention (PCI) procedures. Peri-procedural analysis of ICA images, including automated functional assessment, could optimize PCI outcomes by providing a lesion-specific recommendation on a revascularization strategy, eventually with advice on stent size, length, location, and preferred strategy. Even after stenting, automated measurements on the proportion of stent under expansion and hemodynamic function may inform the operator and patient about the expected short- and long-term outcome.
And finally, we suggest that the comprehensive of requirements of deep learning solutions of ICA images could be potential tools could streamline the calculation of scales to guide clinical decision-making in complex CAD (e.g. the calculation of the SYNTAX score).

The organization of the rest of the paper is structured as follows: Section \ref{sec:methodology} specifies the details of the dataset used and training models. The experimental setup and results are presented in Section \ref{sec:experimental}. Finally, Section \ref{sec:conclusions} is devoted to conclusions.

\section{Methodology} \label{sec:methodology}
\subsection{Source data} \label{sec:sourceData}

The Invasive Coronary Angiography (ICA) dataset is composed of videos from 42 anonymized patients acquired at the Hospital  Universitario Virgen de la Victoria in Málaga (Spain) with Artis Zee (Siemens AG, Muenchen, Germany) as cardiac angiography equipment. They have been included within the regulation set by the local ethical committee of the hospital and patient consent was waived because this is a retrospective study with anonymized data. The dataset includes different projections for the left and right coronary arteries, such as the right and left anterior obliques, with cranial and caudal angulation.

In conjunction with a team of cardiologists, a selection of frames, where the radiocontrast had been perfusing correctly or the lesion was discernible, was done for each video. Furthermore, these frames were annotated, delimiting the region of interest by bounding boxes and organizing them into categories. The possible clinical categories were established according to seven possible lesion degree ranges depending on the narrowing of the vessel, in ascending order: $<20\%$, $[20\%, 49\%]$, $[50\%, 69\%]$, $[70\%, 89\%]$, $[90\%, 98\%]$, $99\%$ and $100\%$. The 99\% and 100\% lesion categories have a particular morphology. A 100\% lesion is a total occlusion of the vessel, from which the continuation of the vessel is imperceptible. The 99\% lesions present a gap, the radiocontrast is imperceptible in the narrow, but the continuation of the vessel is visible. The rest of the categories are assessed depending on the grade of narrowing with respect to the lumen on the vessel. In total, there are 3,900 images with at least one lesion and 1,943 images with no visible lesions.

\subsection{Data preprocessing} \label{sec:dataPre}

\begin{figure*}[ht]
	\centering
	\subfigure[Non-lesion]{\includegraphics[width = 2.5 cm]{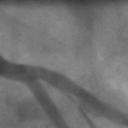}\label{fig:exampleLCA1}} 
        \subfigure[$<$20\%]{\includegraphics[width = 2.5 cm]{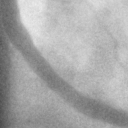}\label{fig:exampleLCA2}} 
        \subfigure[{[20\%, 49\%]}]{\includegraphics[width = 2.5 cm]{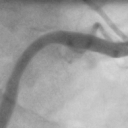}\label{fig:exampleLCA3}} 
        \subfigure[{[50\%, 69\%]}]{\includegraphics[width = 2.5 cm]{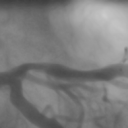}\label{fig:exampleLCA4}} 
        
        \subfigure[{[70\%, 89\%]}]{\includegraphics[width = 2.5 cm]{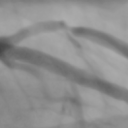}\label{fig:exampleLCA5}}
        \subfigure[{[90\%, 98\%]}]{\includegraphics[width = 2.5 cm]{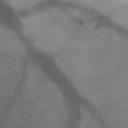}\label{fig:exampleLCA6}} 
        \subfigure[99\%]{\includegraphics[width = 2.5 cm]{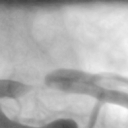}\label{fig:exampleLCA7}} 
        \subfigure[100\%]{\includegraphics[width = 2.5 cm]{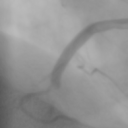}\label{fig:exampleLCA8}}
        
    \caption{Samples of the eight lesion ranges in which patches are categorized.} \label{fig:categories}
\end{figure*}

The present work focuses on the classification of patches, i.e. equal subdivisions, of ICA images. The raw images have size 512 $\times$ 512 pixels, which were divided into a 4 $\times$ 4 grid, and then resized to 32 $\times$ 32 pixels. This way we want to preserve spatial information near the lesion and the downscale was done for training performance. These patches were labeled with the corresponding lesion degree if the centroid of the lesion bounding box falls into it, and the rest as ``non-lesion'' patches. \figurename \, \ref{fig:categories} shows representative samples of patches for each possible category. 

This procedure implied a vast increase in the ``non-lesion'' class, the negative class. To relieve this imbalance between patches with and without lesions, the latter set was reduced before the training. First, the background ``non-lesion'' patches were removed: a basic mask of the ICA images was extracted using morphological operations to segment the vessels to carry out this filtering. The masks were split into patches and those patches where the mask had less than 2\% of vessel pixels were discarded. However, both sets still are unbalanced, so to prevent this issue, a random reduction of the ``non-lesion'' class was applied, equalizing both classes to have the same number of elements.

\begin{figure}[ht]
	\centering
	\subfigure[Original]{\includegraphics[width = 2.5 cm]{P99-LCA_p30_v2_00021_03.png}\label{fig:ori}} 
        \subfigure[Translation]{\includegraphics[width = 2.5 cm]{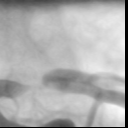}\label{fig:tras}}         
        \subfigure[Scale]{\includegraphics[width = 2.5 cm]{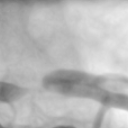}\label{fig:sca}} 
        \subfigure[Flip]{\includegraphics[width = 2.5 cm]{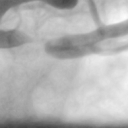}\label{fig:flip}} 
        \label{fig:datagumentation}
        
    \caption{Examples of the modifications applied to the training
sets to augment data.}
\end{figure}

Once both classes had been equalized, data augmentation was employed by applying different random basic spatial operations of the original patches. Particularly, these basic operations were:
\begin{itemize}
    \item Translations in the $X$ and $Y$ axis in a random range of $[-4,4]$ pixels, \figurename \,  \ref{fig:tras}.
    \item Scaling of the images randomly with a scale factor in a
range of $[0.9,1.2]$, \figurename \,  \ref{fig:sca}.
    \item Flip horizontally and vertically, \figurename \,  \ref{fig:flip}.
\end{itemize}

\subsection{Convolutional Neural Networks} \label{sec:cnn} 


In this study, different well-known Convolutional Neural Network (CNN) architectures were employed to analyze their performance concerning the positive class assigned to binary classify ICA images into ``lesion'' and ``non-lesion'' classes. CNNs are based on convolutional layers, where former layers extract basic features, latter layers extract more specific features, and pooling layers are used to subsample features maps, and fully connected layers as final classifier \cite{rawat2017deep,wang2021review}. Five widely used in the literature pre-trained architectures were selected:

\begin{itemize}
    \item DenseNet-201, characterized by implementing dense blocks connecting their layers to all former layers \cite{huang2017densely}.
    \item MobileNet-V2 is a mobile neural network based on the combination of depthwise convolution, which applies to the input a single filter without new features, and pointwise convolution, which produces a linear combination output extracting features \cite{sandler2018mobilenetv2,sae2019convolutional}.
    \item NasNet-Mobile is the smallest model of NasNet versions, whose architectures are designed by Neural Architecture Search (NAS) which finds the best cells or basic blocks using the reinforcement learning technique \cite{reddy2018comparison}.
    \item ResNet-18 and ResNet-50 are Residual Networks (ResNets) that introduce the concept of residual connections, implementing shortcut connections where certain convolutional layers can be skipped at one time \cite{he2016deep,wang2021review}.
\end{itemize}

\subsection{Evaluation metrics} \label{sec:metrics}


In order to quantify the performance of different methods to classify ICA images as a binary classification task, the main four representative parameters are used: True Positive (\textit{TP}), True Negative (\textit{TN}), False Positive (\textit{FP}) and False Negative (\textit{FN}) \cite{ovalle2022hybrid}. F-measure is one of the related metrics that provides good overall performance because it integrates Precision and Recall measures under the concept of harmonic mean \cite{grandini2020metrics}. Precision indicates the rate of correctly positive samples over total positive predicted samples, while Recall is the proportion of correctly positive samples overall actual positive samples. Another measure involving two measures is the area under a ROC curve (AUC), which is also calculated. AUC corresponds to the integral of a ROC curve which shows the Recall versus the Specificity for different thresholds of classification scores. Specificity is the proportion of correctly classified negative samples out of the total actual negative samples. 

The mentioned measures range in $[0,1]$ (the higher is better), and are defined as follows:

\begin{equation}
F\text{-}measure = 2 \cdot \frac{Pre \cdot Rec}{Pre + Rec}
\quad
Precision = \frac{TP}{TP + FP}
\end{equation}
\begin{equation}
Recall = \frac{TP}{TP + FN}
\qquad
Specificity = \frac{TN}{FP + TN}
\end{equation}

\section{Experimental results} \label{sec:experimental} 

\subsection{Training and experiments description}

This work aims to analyze the impact on the performance of the binary ``lesion''/ ``non-lesion'' classification when different degrees of lesions are considered into the positive class. For each experiment, the positive class, i.e., the ``lesion'' class, was set up by lesion degrees, including all higher degrees. For example, $\geq 90\%$ positive class includes $100\%$, $99\%$, and $[98\%,90\%]$ categories. Besides, the ``non-lesion'' class was randomly reduced to equalize the number of patches of the lesion class, as above mentioned. Both classes were divided into training (80\%) and test (20\%) sets by videos, i.e., frames of the same video in the train set are unavailable for the test set because frames of the same sequence are very similar. This way allows for estimating a fairer performance evaluation.

Due to the clear morphological difference between lesions of 100\% and 99\% severity compared to the remaining degrees, four training strategies were established, including or not these high levels of severity and the use of data augmentation. The first experiment, named ``\textit{With 100\% and 99\% lesions}'', contemplates seven categories, i. e. all possible degrees, therefore seven positive classes are established, in each of which the higher former degrees are included. On the contrary, another strategy, ``\textit{W/o 100\% and 99\% lesions}'', considers only five categories, excluding lesions of 100\% and 99\% severity, so five positive classes are determined. Finally, two extra strategies were considered by applying data augmentation to the training set (80\% of each class), doubling the amount of data. The number of patches used in the training process is reported in \tablename \, \ref{tab:numpatches}. It must be considered that the sizes of train sets of adjacent categories may mismatch because of the different number of frames in video sequences.

\begin{table*}[ht]
    \centering
    \caption{Number of patches used as the training set for each established strategy.}
    \begin{tabular}{|c|c|c|c|c|}
    \hline
      & \makecell{With 100\% and \\ 99\% lesions} &  \makecell{With 100\% and \\ 99\% lesions + Data \\ Augmentation} & \makecell{W/o 100\% and \\ 99\% lesions} & \makecell{W/o 100\% and \\ 99\% lesions + Data \\ Augmentation} \\                                         
    \hline
    100\% & 242 & 484 & - & - \\
    $\geq 99\%$ & 258 & 516& - & - \\
    $\geq 90\%$ & 874 & 1748 & 706 & 1412 \\
    $\geq 70\%$ & 1442 & 2884 & 1386 & 2772 \\
    $\geq 50\%$ & 3170 & 6340 & 2820 & 5640\\
    $\geq 20\%$ & 5064 & 10128 & 4894 & 9788\\
    $>$ 0\% & 8626 & 17252 & 8230 & 16460\\
    \hline
    \end{tabular}
    \label{tab:numpatches}
\end{table*}

Regarding the CNNs training, we set some hyperparameters: validation frequency = 50, validation patience = 5, and maximum epochs = 50, while the batch size was set according to the number of training patches to keep the rate of iterations in all training processes. In contrast, we tuned up the optimizer and the initial learning rate. Three different algorithms were compared: Adam (adaptive moment estimation), SGDM (Stochastic Gradient Descent with Momentum), and RMSProp (Root Mean Square Propagation). Four initial learning rates were tested: 0.01, 0.001, 0.0001, and 0.00001. In total, there are 12 possible hyperparameter combinations for each threshold delimiting the positive class in each strategy and network. 5-fold stratified cross-validation was implemented to compare all these possibilities reliably. The learning rate and optimizer were selected for each CNN based on the average validation accuracy among the 5 folds.

The proposed models were implemented in MATLAB R2022b on a computer system with an Intel Core i9-10900X processor, 128 GB of RAM, and NVIDIA GeForce RTX 3080 Ti GPU card. Furthermore, no layer of chosen pre-trained methods was frozen, so all weights were updated during the training process according to the input class information.

\subsection{Results} \label{sec:results}

\begin{figure*}[t]
    \centering
    \includegraphics[width=\textwidth]{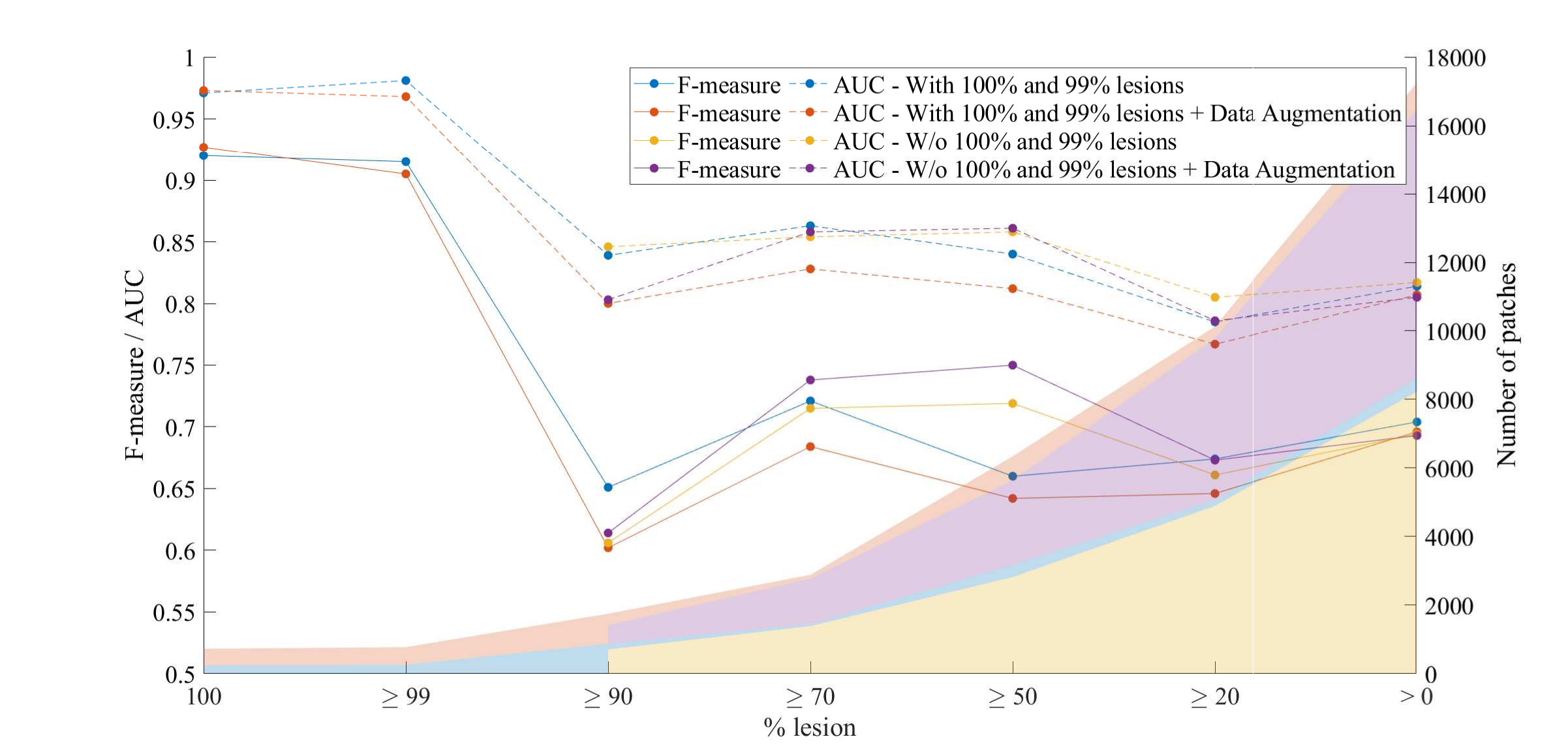}
    \caption{Highest F-measure and AUC obtained in the test set with 5-fold cross-validation and the number of patches used as the train set.}
    \label{fig:finalSelection}
\end{figure*}

Next, we present the result outcomes of the experiments above, based on applying the optimized model, and evaluating the validation accuracy obtained in the training process, to the test set. In \tablename \, \ref{tab:1stSel-F} and \tablename \, \ref{tab:1stSel-AUC}, the F-measure and AUC of the chosen model are reported, respectively. The highest F-measure and AUC values by positive class and strategy established are shown in bold, standing out one architecture among the 5 CNNs for each case. These highest values are plotted in \figurename \,  \ref{fig:finalSelection}, where the numbers of patches for the training process are depicted too.

\begin{table*}[ht]
    \centering
    \caption{F-measure obtained on the test set using 5-fold stratified cross-validation for the four strategies. The highest values by rows are shown in bold.}
    \resizebox{\textwidth}{!}{
    \begin{tabular}{|c|c|ccccc|}
    \hline

    \multirow{2}{*}{Strategy} &\multirow{2}{*}{\shortstack{Lesion \\ Range}} & \multicolumn{5}{c|}{Convolutional Neural Networks Models} \\ 
                                                       
    & & DenseNet-201 & MobileNet-V2 & NasNet-Mobile & ResNet-18 & ResNet-50 \\
    \hline
    \multirow{7}{*}{\shortstack{With \\ 100\% and \\ 99\% lesions}} 
    & 100\% & $0.897 \pm 0.019$ & $0.866 \pm 0.091$ & $0.884 \pm 0.056$ & $0.853 \pm 0.089$ & $\boldsymbol{0.920 \pm 0.031}$ \\
    & $\geq 99\%$ & $\boldsymbol{0.915 \pm 0.026}$ & $0.915 \pm 0.040$ & $0.908 \pm 0.059$ & $0.850 \pm 0.110$ & $0.893 \pm 0.093$ \\
    & $\geq 90\%$ & $0.592 \pm 0.082$ & $0.575 \pm 0.031$ & $0.574 \pm 0.060$ & $\boldsymbol{0.651 \pm 0.051}$ & $0.582 \pm 0.047$ \\
    & $\geq 70\%$ & $\boldsymbol{0.721 \pm 0.049}$ & $0.635 \pm 0.057$ & $0.655 \pm 0.035$ & $0.677 \pm 0.040$ & $0.654 \pm 0.019$ \\
    & $\geq 50\%$ & $\boldsymbol{0.660 \pm 0.027}$ & $0.647 \pm 0.035$ & $0.632 \pm 0.045$ & $0.631 \pm 0.057$ & $0.624 \pm 0.023$ \\
    & $\geq 20\%$ & $\boldsymbol{0.674 \pm 0.025}$ & $0.669 \pm 0.046$ & $0.655 \pm 0.051$ & $0.651 \pm 0.028$ & $0.633 \pm 0.036$ \\
    & $>$ 0\% & $0.690 \pm 0.019$ & $\boldsymbol{0.704 \pm 0.012}$ & $0.673 \pm 0.030$ & $0.679 \pm 0.029$ & $0.675 \pm 0.019$ \\
    \hline
    \multirow{7}{*}{\shortstack{With \\ 100\% and \\ 99\% lesions \\+ Data \\ Augmentation}} 
    & 100\% & $0.894 \pm 0.034$ & $0.893 \pm 0.098$ & $0.803 \pm 0.081$ & $\boldsymbol{0.927 \pm 0.032}$ & $0.914 \pm 0.060$ \\
    & $\geq 99\%$ & $\boldsymbol{0.905 \pm 0.023}$ & $0.860 \pm 0.040$ & $0.894 \pm 0.038$ & $0.868 \pm 0.045$ & $0.884 \pm 0.017$ \\
    & $\geq 90\%$ & $0.520 \pm 0.062$ & $0.522 \pm 0.118$ & $0.479 \pm 0.069$ & $0.582 \pm 0.101$ & $\boldsymbol{0.602 \pm 0.033}$ \\
    & $\geq 70\%$ & $0.676 \pm 0.035$ & $0.683 \pm 0.024$ & $\boldsymbol{0.684 \pm 0.024}$ & $0.652 \pm 0.081$ & $0.637 \pm 0.055$ \\
    & $\geq 50\%$ & $0.600 \pm 0.056$ & $0.636 \pm 0.062$ & $\boldsymbol{0.642 \pm 0.053}$ & $0.619 \pm 0.079$ & $0.540 \pm 0.029$ \\
    & $\geq 20\%$ & $0.645 \pm 0.019$ & $\boldsymbol{0.646 \pm 0.009}$ & $0.633 \pm 0.029$ & $0.644 \pm 0.017$ & $0.615 \pm 0.024$ \\
    & $>$ 0\% & $0.689 \pm 0.064$ & $0.672 \pm 0.030$ & $\boldsymbol{0.696 \pm 0.020}$ & $0.693 \pm 0.009$ & $0.678 \pm 0.026$ \\
    \hline
    \multirow{5}{*}{\shortstack{W/o \\ 100\% and \\ 99\% lesions}}
    & $\geq 90\%$ & $0.596 \pm 0.148$ & $0.497 \pm 0.119$ & $\boldsymbol{0.606 \pm 0.106}$ & $0.561 \pm 0.123$ & $0.420 \pm 0.070$ \\
    & $\geq 70\%$ & $\boldsymbol{0.715 \pm 0.053}$ & $0.701 \pm 0.082$ & $0.666 \pm 0.016$ & $0.659 \pm 0.124$ & $0.598 \pm 0.069$ \\
    & $\geq 50\%$ & $0.711 \pm 0.047$ & $\boldsymbol{0.719 \pm 0.034}$ & $0.694 \pm 0.015$ & $0.700 \pm 0.067$ & $0.673 \pm 0.079$ \\
    & $\geq 20\%$ & $0.659 \pm 0.024$ & $0.641 \pm 0.022$ & $0.657 \pm 0.028$ & $\boldsymbol{0.661 \pm 0.030}$ & $0.605 \pm 0.069$ \\
    & $>$ 0\% & $\boldsymbol{0.694 \pm 0.031}$ & $0.666 \pm 0.015$ & $0.669 \pm 0.019$ & $0.658 \pm 0.031$ & $0.656 \pm 0.032$ \\
    \hline
    \multirow{5}{*}{\shortstack{W/o \\ 100\% and \\ 99\% lesions \\+ Data \\ Augmentation}}
    & $\geq 90\%$ & $0.494 \pm 0.057$ & $\boldsymbol{0.614 \pm 0.118}$ & $0.558 \pm 0.103$ & $0.450 \pm 0.055$ & $0.446 \pm 0.115$ \\
    & $\geq 70\%$ & $\boldsymbol{0.738 \pm 0.052}$ & $0.707 \pm 0.061$ & $0.721 \pm 0.044$ & $0.671 \pm 0.065$ & $0.567 \pm 0.025$ \\
    & $\geq 50\%$ & $\boldsymbol{0.750 \pm 0.052}$ & $0.740 \pm 0.034$ & $0.729 \pm 0.007$ & $0.695 \pm 0.026$ & $0.675 \pm 0.032$ \\
    & $\geq 20\%$ & $0.651 \pm 0.032$ & $0.655 \pm 0.031$ & $\boldsymbol{0.673 \pm 0.022}$ & $0.660 \pm 0.020$ & $0.664 \pm 0.018$ \\
    & $>$ 0\% & $0.686 \pm 0.029$ & $0.676 \pm 0.012$ & $\boldsymbol{0.693 \pm 0.008}$ & $0.663 \pm 0.028$ & $0.676 \pm 0.024$ \\
    \hline
    \end{tabular}    
    } \label{tab:1stSel-F}
\end{table*}

\begin{table*}[ht]
    \centering
    \caption{AUC obtained on the test set using 5-fold stratified cross-validation, for the four strategies. The highest values by rows are shown in bold.}
    \resizebox{\textwidth}{!}{
    \begin{tabular}{|c|c|ccccc|}
    \hline

    \multirow{2}{*}{Strategy} &\multirow{2}{*}{\shortstack{Lesion \\ Range}} & \multicolumn{5}{c|}{Convolutional Neural Networks Models} \\ 
                                                       
    & & DenseNet-201 & MobileNet-V2 & NasNet-Mobile & ResNet-18 & ResNet-50 \\
    \hline
    \multirow{7}{*}{\shortstack{With \\ 100\% and \\ 99\% lesions}} 
    & 100\% & $\boldsymbol{0.971 \pm 0.023}$ & $0.950 \pm 0.035$ & $0.943 \pm 0.039$ & $0.969 \pm 0.020$ & $0.958 \pm 0.013$ \\
    & $\geq 99\%$ & $\boldsymbol{0.981 \pm 0.013}$ & $0.967 \pm 0.027$ & $0.952 \pm 0.075$ & $0.954 \pm 0.028$ & $0.967 \pm 0.045$ \\
    & $\geq 90\%$ & $0.800 \pm 0.024$ & $0.746 \pm 0.029$ & $0.800 \pm 0.063$ & $\boldsymbol{0.839 \pm 0.026}$ & $0.825 \pm 0.030$ \\
    & $\geq 70\%$ & $\boldsymbol{0.863 \pm 0.024}$ & $0.845 \pm 0.028$ & $0.823 \pm 0.014$ & $0.837 \pm 0.021$ & $0.823 \pm 0.015$ \\
    & $\geq 50\%$ & $\boldsymbol{0.840 \pm 0.011}$ & $0.791 \pm 0.019$ & $0.828 \pm 0.028$ & $0.815 \pm 0.038$ & $0.786 \pm 0.046$ \\
    & $\geq 20\%$ & $\boldsymbol{0.785 \pm 0.009}$ & $0.775 \pm 0.010$ & $0.779 \pm 0.021$ & $0.783 \pm 0.011$ & $0.767 \pm 0.012$ \\
    & $>$ 0\% & $0.807 \pm 0.009$ & $0.813 \pm 0.004$ & $\boldsymbol{0.814 \pm 0.012}$ & $0.806 \pm 0.007$ & $0.807 \pm 0.013$ \\
    \hline
    \multirow{7}{*}{\shortstack{With \\ 100\% and \\ 99\% lesions \\+ Data \\ Augmentation}} 
    & 100\% & $0.949 \pm 0.022$ & $0.924 \pm 0.073$ & $0.914 \pm 0.052$ & $\boldsymbol{0.973 \pm 0.017}$ & $0.973 \pm 0.026$ \\
    & $\geq 99\%$ & $\boldsymbol{0.968 \pm 0.017}$ & $0.930 \pm 0.025$ & $0.954 \pm 0.027$ & $0.944 \pm 0.005$ & $0.952 \pm 0.029$ \\
    & $\geq 90\%$ & $0.732 \pm 0.046$ & $0.761 \pm 0.035$ & $0.749 \pm 0.035$ & $0.749 \pm 0.041$ & $\boldsymbol{0.800 \pm 0.017}$ \\
    & $\geq 70\%$ & $\boldsymbol{0.828 \pm 0.022}$ & $0.809 \pm 0.030$ & $0.825 \pm 0.016$ & $0.820 \pm 0.041$ & $0.783 \pm 0.031$ \\
    & $\geq 50\%$ & $0.799 \pm 0.037$ & $0.782 \pm 0.032$ & $\boldsymbol{0.812 \pm 0.022}$ & $0.795 \pm 0.020$ & $0.743 \pm 0.016$ \\
    & $\geq 20\%$ & $\boldsymbol{0.767 \pm 0.022}$ & $0.753 \pm 0.016$ & $0.751 \pm 0.021$ & $0.749 \pm 0.021$ & $0.739 \pm 0.019$ \\
    & $>$ 0\% & $\boldsymbol{0.807 \pm 0.042}$ & $0.797 \pm 0.013$ & $0.796 \pm 0.024$ & $0.807 \pm 0.005$ & $0.785 \pm 0.012$ \\
    \hline
    \multirow{5}{*}{\shortstack{W/o \\ 100\% and \\ 99\% lesions}}
    & $\geq 90\%$ & $\boldsymbol{0.846 \pm 0.040}$ & $0.771 \pm 0.051$ & $0.800 \pm 0.047$ & $0.828 \pm 0.047$ & $0.772 \pm 0.071$ \\
    & $\geq 70\%$ & $\boldsymbol{0.854 \pm 0.010}$ & $0.818 \pm 0.061$ & $0.838 \pm 0.013$ & $0.823 \pm 0.060$ & $0.800 \pm 0.021$ \\
    & $\geq 50\%$ & $\boldsymbol{0.858 \pm 0.015}$ & $0.853 \pm 0.020$ & $0.820 \pm 0.012$ & $0.844 \pm 0.026$ & $0.829 \pm 0.029$ \\
    & $\geq 20\%$ & $\boldsymbol{0.805 \pm 0.014}$ & $0.785 \pm 0.018$ & $0.772 \pm 0.014$ & $0.794 \pm 0.027$ & $0.765 \pm 0.054$ \\
    & $>$ 0\% & $\boldsymbol{0.817 \pm 0.014}$ & $0.789 \pm 0.012$ & $0.794 \pm 0.011$ & $0.801 \pm 0.016$ & $0.789 \pm 0.026$ \\
    \hline
    \multirow{5}{*}{\shortstack{W/o \\ 100\% and \\ 99\% lesions \\+ Data \\ Augmentation}}
    & $\geq 90\%$ & $0.777 \pm 0.041$ & $0.794 \pm 0.061$ & $\boldsymbol{0.803 \pm 0.035}$ & $0.729 \pm 0.052$ & $0.727 \pm 0.066$ \\
    & $\geq 70\%$ & $\boldsymbol{0.858 \pm 0.013}$ & $0.790 \pm 0.031$ & $0.828 \pm 0.025$ & $0.805 \pm 0.024$ & $0.760 \pm 0.011$ \\
    & $\geq 50\%$ & $\boldsymbol{0.861 \pm 0.026}$ & $0.830 \pm 0.018$ & $0.838 \pm 0.014$ & $0.830 \pm 0.023$ & $0.825 \pm 0.020$ \\
    & $\geq 20\%$ & $0.780 \pm 0.021$ & $0.772 \pm 0.025$ & $0.778 \pm 0.008$ & $\boldsymbol{0.786 \pm 0.014}$ & $0.771 \pm 0.028$ \\
    & $>$ 0\% & $\boldsymbol{0.805 \pm 0.014}$ & $0.776 \pm 0.011$ & $0.788 \pm 0.015$ & $0.780 \pm 0.014$ & $0.777 \pm 0.012$ \\
    \hline
    \end{tabular}
    } \label{tab:1stSel-AUC}
\end{table*}

By analyzing \tablename \, \ref{tab:1stSel-F}, the first remarkable aspect is that all lesion ranges established as the positive class have the same tendencies independently of the strategy or methods employed. Note that the best results are attained with high-severe lesions; for 100\% positive class, ResNet-18 with data augmentation and ResNet-50 stand out with the highest values, 0.927 and 0.920, respectively. However, for $\geq 99\%$, the DenseNet-201 model obtained the best outcomes for both with and without data augmentation. DenseNet-201 achieved good results even considering lower degrees ($\geq 70\%$, $\geq 50\%$ and $\geq 20\%$) but its performance decreased sightly when data augmentation is applied, being NasNet-Mobile more robust. If 100\% and 99\% lesions are excluded, no model stands out above the rest. For instance, DenseNet-201 and MobileNet-V2 with data augmentation have a fair-to-high performance for $\geq 70\%$ and $\geq 50\%$ positive class, around 0.7, being the rest of the models under this. Finally, ResNet-50 for $\geq 90\%$ positive class yielded a poor result, below 50\% of F-measure.

Regarding \tablename \, \ref{tab:1stSel-AUC}, AUC values are reported. The AUC measure considers the specificity and the recall, which helps to check how well each class is classified. The first outstanding fact is that any value is below 0.7, the lowest value is 0.727 for $\geq 90\%$ and using ResNet-50 with the ``W/o 100\% and 99\% lesions + Data augmentation'' strategy. It is a fair-to-high value but the corresponding F-measure from \tablename \, \ref{tab:1stSel-F} is a poor value, 0.446. This fact points out that AUC needs to be supported by another performance metric. As F-measure, all architectures obtain similar results along all positive classes and strategies. The highest values are obtained with high-severe lesion ranges: 0.981 with DenseNet-201 for $\geq 99\%$ without data augmentation, 0.973 with ResNet-18 and data augmentation, and 0.971 got with DenseNet-201, both for 100\% positive class. In this case, the outcomes obtained decrease slightly when moderate and mild lesion degrees are considered into the positive class. This fact makes sense because lesions are more complex to discern from healthy vessels. For positive classes which include lower lesion degrees, AUC decreases under 0.9. For $\geq 90\%$, being the highest values: 0.863 for $\geq 70\%$, and 0.861 for $\geq 50\%$, using ``With 100\% and 99\% lesions'' and ``W/o 100\% and 99\% lesions + Data augmentation'' strategies, respectively, and both with DenseNet-201. Considering AUC, DenseNet-201 is clearly the architecture that stands out over the rest, achieving the highest results in most cases.


Focusing on \figurename \, \ref{fig:finalSelection}, this graph shows the tendencies for the four implemented strategies. The behavior of the performance of the strategies is similar along all ``lesion'' classes independently of the strategy followed. It can be seen clearly how the positive classes $100\%$ and $\geq 99\%$  are very well classified, either considering F-measure or AUC, in spite of the small number of patches used. Whereas when the $[98\%,90\%]$ range is included the performance drops significantly, around 15\% and 25\% for F-measure and AUC, respectively. The lower outcomes attained with the $\geq 90\%$ category are similar along all strategies. Then, the performance increases for $\geq 70\%$ and $\geq 50\%$ positive classes, decreasing slightly again when the lowest range degrees are included. This tendency is followed by both measures, F-measure and AUC. It could be interpreted considering the kind of lesions and the number of patches. Despite the growth of the number of ``lesion'' patches, the results do not improve because the classification task becomes more complex, as lower lesion degrees are more difficult to discern from ``non-lesion'' patches. Therefore, categories of 100\% and $\geq 99\%$ lesions are well classified, achieving excellent results because of their clear morphological difference, despite the small number of patches used. Furthermore, $\geq 70\%$ and $\geq 50\%$ ranges have good results because a larger number of patches are employed, and the lesions considered remain clearly distinguishable. Also, it could stand out that the augmentation data implies null improvement when 100\% and 99\% lesion degrees are considered and a slight improvement when they are excluded. This fact supports the idea that these fine-grain categories represent a highly complex problem, since despite the data augmentation applied, the methods still have difficulties in improving their performance. Additionally, there is a great difference between AUC and F-measure values, both measures involve Recall, which measures how well is classified the positive class (``lesion'' class in this case), but AUC takes into account Specificity, the rate of how well classifies is the negative class, and F-measure considers Precision, which rates the positive samples correctly classified. Considering the difference obtained between measures, it could be interpreted as better Specificity than Precision, which means that the negative class, i.e. the ``non-lesion'' class, in some cases is slightly better classified than the positive class.

\begin{figure*}[ht]
	\centering
	\subfigure[Strategies including 100\% and 99\% lesion degrees.]{\includegraphics[width=0.49\textwidth]{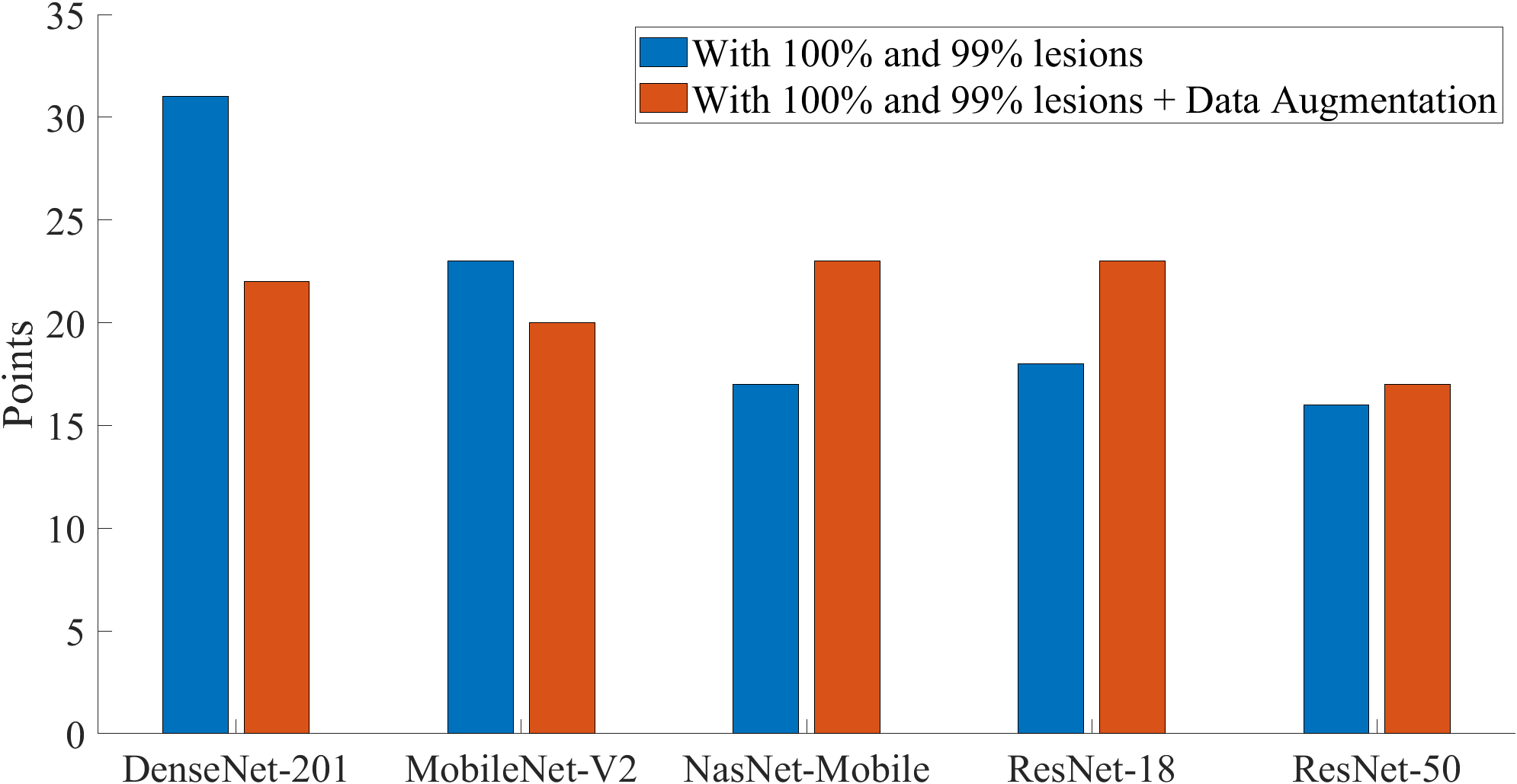}\label{fig:rankingW-F}} 
        \subfigure[Strategies excluding 100\% and 99\% lesion degrees.]{\includegraphics[width=0.49\textwidth]{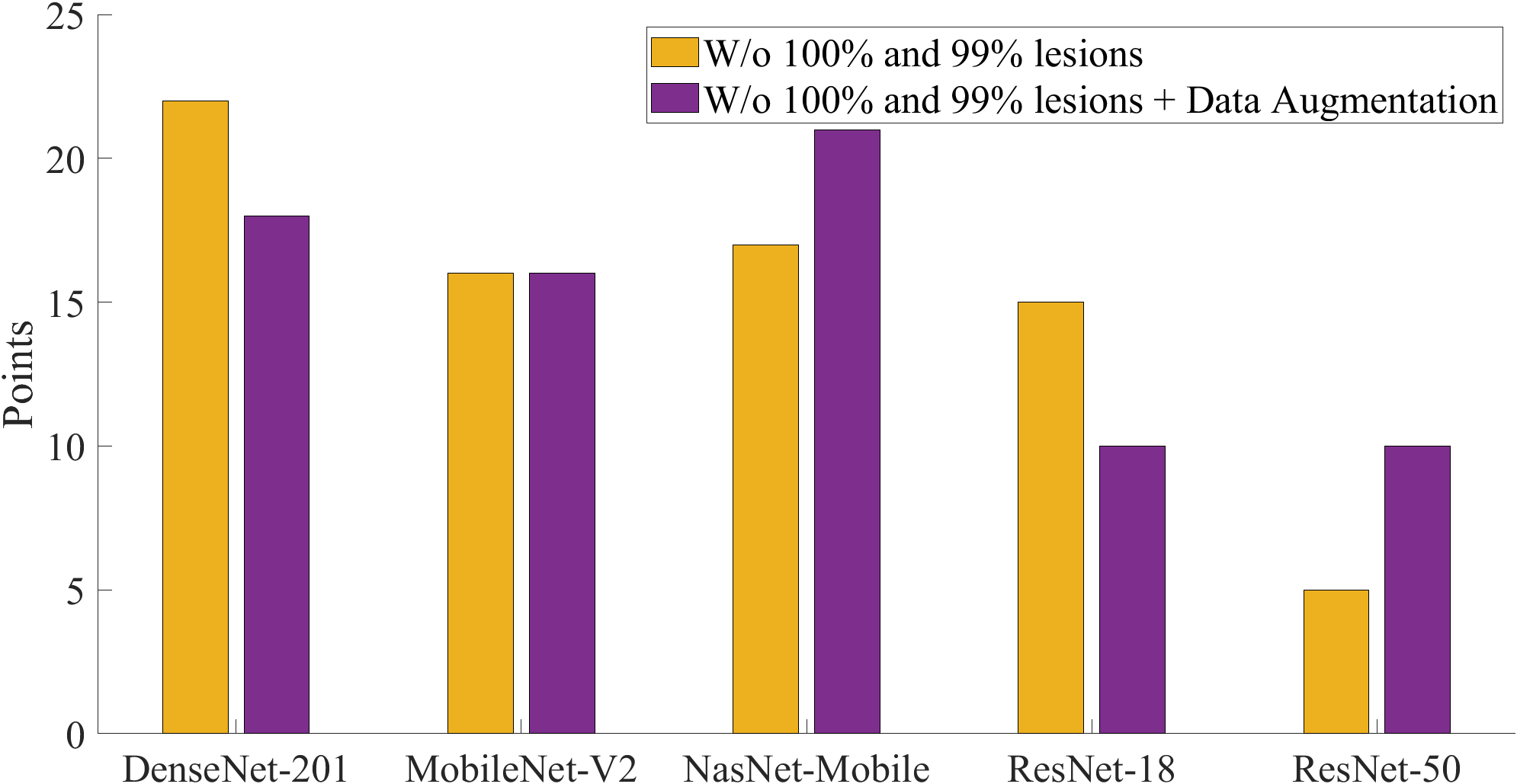}\label{fig:rankingN-F}}        
\caption{Ranking of methods considering F-measure obtained for each positive class established.}\label{fig:rankingF}

        \subfigure[Strategies including 100\% and 99\% lesion degrees.]{\includegraphics[width=0.49\textwidth]{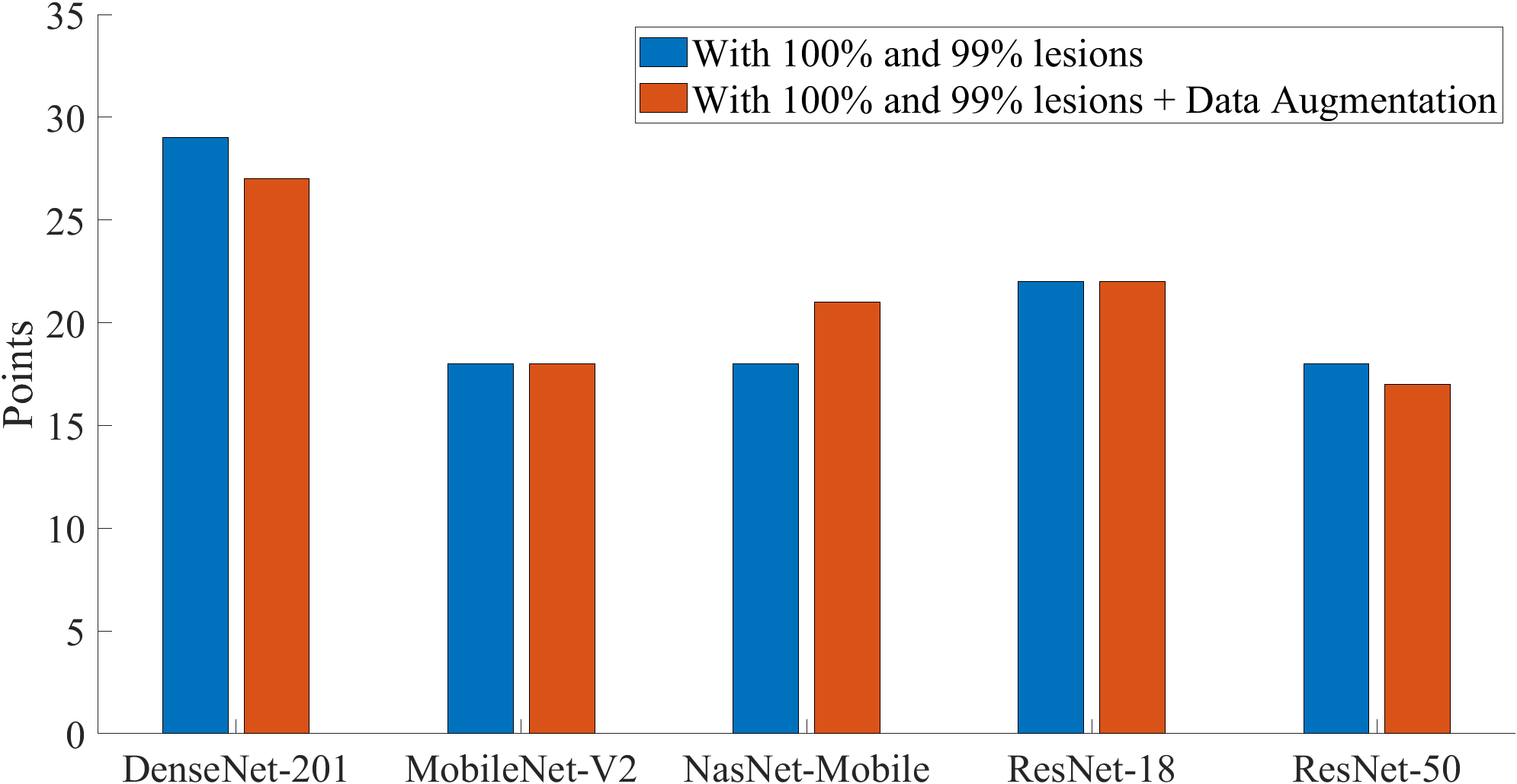}\label{fig:rankingW-AUC}} 
        \subfigure[Strategies excluding 100\% and 99\% lesion degrees.]{\includegraphics[width=0.49\textwidth]{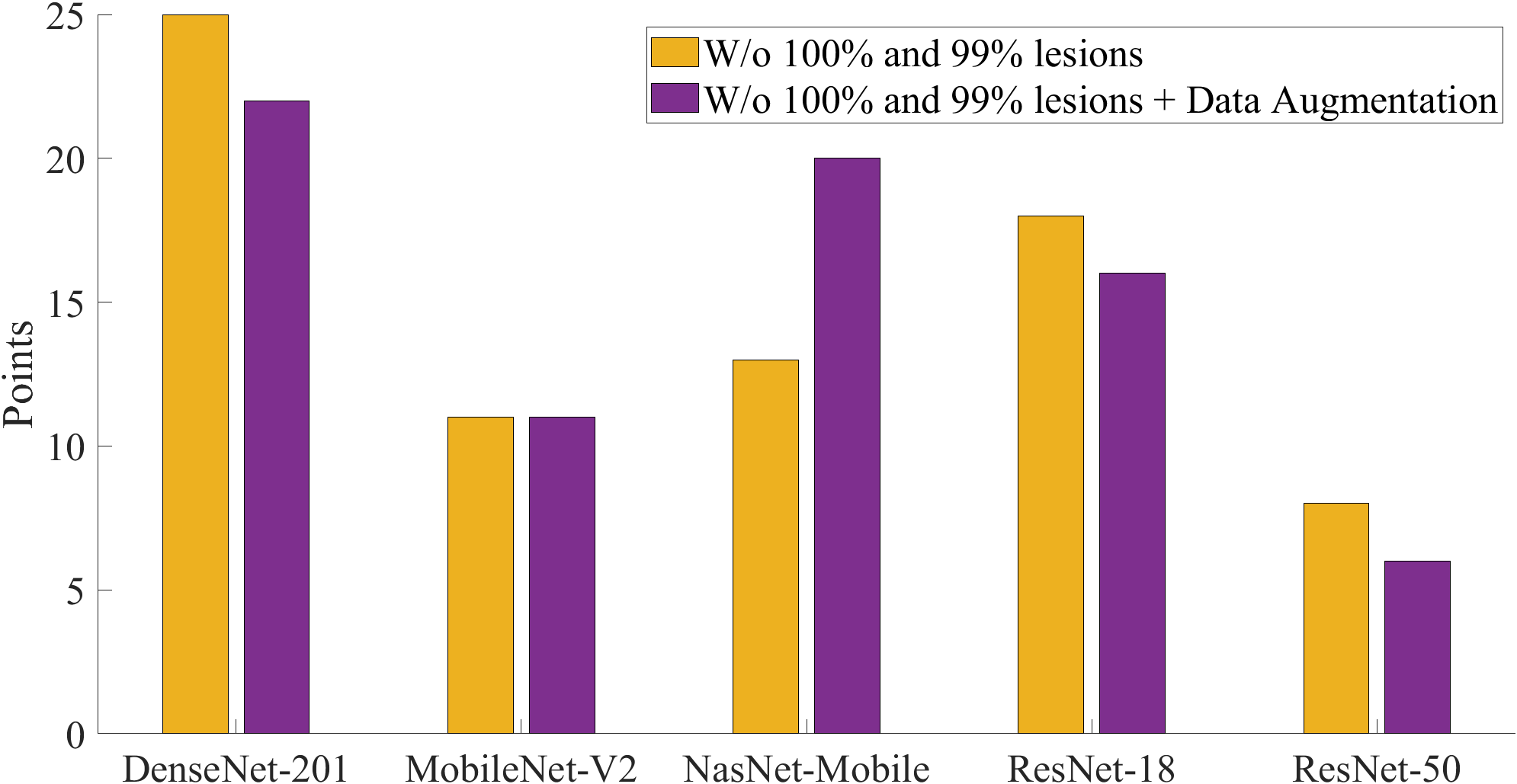}\label{fig:rankingN-AUC}}
        
    \caption{Ranking of methods considering AUC obtained for each positive class established.}\label{fig:rankingAUC}
\end{figure*}

 In addition, a ranking among the proposed architectures was computed in \figurename s\  \ref{fig:rankingF} and \ref{fig:rankingAUC}, where obtained points are divided by strategies. The scores were set by sorting the corresponding performance metric, F-measure or AUC, obtained by positive class in ascending order, considering better a higher value. The position indicates the points obtained. The points obtained for each lesion range were accumulated for each model. There are 7 and 5 lesion ranges, resp., including or excluding 100\% and 99\%, and 5 methods, so the maximum possible score is 35 and 25 points, respectively.  Focusing on strategies without data augmentation, DenseNet-201 got the highest points for the four cases, while when data augmentation is applied, points are more spread out, standing out DenseNet and NasNet-Mobile. Achieving the highest points means the model is suitable to solve most of the binary classification problems analyzed here with good performance. Then, despite the type of strategy followed, DenseNet-201 stands out as the best classification network for CAD lesions.

The test sets have different sizes since positive classes are cumulative, i.e., they grow. Therefore, a more restrictive test was carried out, where all established problems use the same test size. To do it, test sets were randomly reduced to the lowest number of patches of the test sets, in this case, 62 patches in the 100\% lesion category. The results obtained are provided in the Supplementary Material. In \tablename \, \ref{tab:test31-dense} are reported the computed F-measure and AUC obtained with DenseNet-201, one of the most suitable architectures according to the results obtained above for F-measure, AUC, and ranking evaluation. Concerning \tablename \, \ref{tab:test31-dense}, the assumptions made previously are also corroborated in this case, despite the reduction of the test set. 100\% and $\geq 99\%$, with and without data augmentation, achieve the highest values ($>85\%$ F-measure, $>95\%$ AUC), suffering a tough decrease when milder categories are included, falling to unacceptable results lower than 50\% of F-score, although AUC remains a fair-to-high value ($>70\%$). In all cases, AUC attains higher values than the F-measure, this fact remarks that the positive class is worse classified than the negative class, despite the class balance applied. Additionally, models trained with data augmentation neither increase their performance substantially, reinforcing the necessity of increasing the lesion data and equalizing categories to avoid specialization on ``non-lesion'' patches.

\begin{table}[ht]
    \centering
    \caption{F-measure and AUC obtained on the test set using 5-fold stratified cross-validation with DenseNet-201 for the four strategies and the same size for the test sets.}
    \begin{tabular}{|c|c|cc|}
    \hline

    Strategy &\shortstack{Lesion Range} & F-measure & AUC\\
                                                       
    \hline
    \multirow{7}{*}{\shortstack{With \\ 100\% and \\ 99\% lesions}} 
    & 100\% & $0.897 \pm 0.019$ & $0.971 \pm 0.023$ \\
    & $\geq 99\%$ & $ \boldsymbol{0.900\pm0.033} $ & $\boldsymbol{0.983\pm 0.011}$ \\
    & $\geq 90\%$ & $0.599 \pm 0.058$& $0.738 \pm 0.044$ \\
    & $\geq 70\%$ & $0.661\pm 0.110$ & $0.834 \pm 0.088$ \\
    & $\geq 50\%$ & $0.534 \pm 0.126$ & $0.796 \pm 0.058$ \\
    & $\geq 20\%$ & $0.706 \pm 0.031$ & $0.820 \pm 0.045$ \\
    & $>$ 0\% & $0.669 \pm 0.088$ & $0.805 \pm 0.052$ \\
    \hline
    \multirow{7}{*}{\shortstack{With \\ 100\% and \\ 99\% lesions \\+ Data \\ Augmentation}} 
    & 100\% & $\boldsymbol{0.894 \pm 0.034}$ & $\boldsymbol{0.949 \pm 0.022}$ \\
    & $\geq 99\%$ & $0.851 \pm 0.157$ & $0.933 \pm 0.092$ \\
    & $\geq 90\%$ & $0.597 \pm 0.066$ & $0.709 \pm 0.034$ \\
    & $\geq 70\%$ & $0.699 \pm 0.024$ & $0.829 \pm 0.020$ \\
    & $\geq 50\%$ & $0.533 \pm 0.103$ & $0.796 \pm 0.033$ \\
    & $\geq 20\%$ & $0.671 \pm 0.025$ & $0.791 \pm 0.030$ \\
    & $>$ 0\% & $0.651 \pm 0.098$ & $0.777 \pm 0.056$ \\
    \hline
    \multirow{5}{*}{\shortstack{W/o \\ 100\% and \\ 99\% lesions}}
    & $\geq 90\%$ & $0.627 \pm 0.171$ & $0.811 \pm 0.034$ \\
    & $\geq 70\%$ & $\boldsymbol{0.781 \pm 0.068}$ & $0.816 \pm 0.058$ \\
    & $\geq 50\%$ & $0.667 \pm 0.050$ & $\boldsymbol{0.843 \pm 0.035}$ \\
    & $\geq 20\%$ & $0.634 \pm 0.055$ & $0.784 \pm 0.023$ \\
    & $>$ 0\% & $0.583 \pm 0.096$ & $0.778 \pm 0.059$ \\
    \hline
    \multirow{5}{*}{\shortstack{W/o \\ 100\% and \\ 99\% lesions \\+ Data \\ Augmentation}}
    & $\geq 90\%$ & $0.462 \pm 0.088$ & $0.713 \pm 0.071$ \\
    & $\geq 70\%$ & $0.650 \pm 0.109$ & $0.773 \pm 0.077$ \\
    & $\geq 50\%$ & $\boldsymbol{0.769 \pm 0.072}$ & $\boldsymbol{0.863 \pm 0.023}$ \\
    & $\geq 20\%$ & $0.608 \pm 0.049$ & $0.772 \pm 0.049$ \\
    & $>$ 0\% & $0.593 \pm 0.042$ & $0.767 \pm 0.036$ \\
    \hline
    \end{tabular}
    \label{tab:test31-dense}
\end{table}

\section{Conclusions} \label{sec:conclusions}

This work presents a classification methodology for coronary artery disease using invasive coronary angiography images. A total of 5 state-of-the-art deep neural models were used to distinguish between lesion and non-lesion images, varying the threshold of lesion degree to consider into the ``lesion'' class. The dataset was divided into non-overlapping patches and four types of experiments were carried out, including data augmentation and removing high-severe classes. 

Results showed that the 99\% and 100\% categories are easy to classify as lesions ($>$90\% F-measure, $>$95\% AUC) even with little data, while when a lower degree is included in the positive class, the performance drops significantly (65\% F-measure, 80\% AUC). If those extreme cases are discarded, the networks reach 75\% of F-measure and 85\% AUC when data augmentation is applied when $\geq 70\%$ and $\geq 50\%$ severity is intended to be detected. Besides, DenseNet-201 and NasNet-Mobile demonstrated their effectiveness in solving most of the binary classification problems raised. 

Further work will be focused on improving the overall classification performance. On one side, classifying each severity degree separately and including more sophisticated preprocessing steps could bring more homogeneity and therefore produce better results. Another approach would be the training of custom deep networks from scratch, using structures that focus one local spatial features.

\section*{Declarations}

\subsection*{Data availability}
CADICA dataset is open-access available at the Mendeley Data repository with the data identification number: 10.17632/p9bpx9ctcv.1, and direct URL to data: \url{https://data.mendeley.com/datasets/p9bpx9ctcv/1}.

\subsection*{Author Contibutions}
All authors listed have made a substantial, direct, and intellectual
contribution to the work, and approved it for publication. 

\subsection*{Declaration of competing interest}
The authors declare that the research was conducted in the absence
of any commercial or financial relationships that could be construed as a potential conflict of interest.

\subsection*{Conflicts of Interest}
The authors declare that they have no conflicts of interest to report regarding the present study

\subsection*{Supplementary information}

Additionally, regarding the experiments with balanced test sets carried out in the end of the experimental section, which are summarized in \tablename \, \ref{tab:test31-dense}, we provide the complete results in Supplementary Material \ref{sec:supplementary}.

\subsection*{Acknowledgment}
This work is partially supported by the Autonomous Government of Andalusia (Spain) under project UMA20-FEDERJA-108, project name Detection, characterization and prognosis value of the non-obstructive coronary disease with deep learning, and also by the Ministry of Science and Innovation of Spain, grant number PID2022-136764OA-I00, project name Automated Detection of Non Lesional Focal Epilepsy by Probabilistic Diffusion Deep Neural Models. It includes funds from the European Regional Development Fund (ERDF). It is also partially supported by the University of M\'alaga (Spain) under grants B1-2019\_01, project name Anomaly detection on roads by moving cameras; B1-2019\_02, project name Self-Organizing Neural Systems for Non-Stationary Environments; B1-2021\_20, project name Detection of coronary stenosis using deep learning applied to coronary angiography; B4-2022, project name Intelligent Clinical Decision Support System for Non-Obstructive Coronary Artery Disease in Coronarographies; B1-2022\_14, project name Detecci\'on de trayectorias an\'omalas de veh\'iculos en c\'amaras de tr\'afico; and by the Fundaci\'on Unicaja under project PUNI-003\_2023, project name Intelligent System to Help the Clinical Diagnosis of Non-Obstructive Coronary Artery Disease in Coronary Angiography. The authors thankfully acknowledge the computer resources, technical expertise and assistance provided by the SCBI (Supercomputing and Bioinformatics) center of the University of M\'alaga. They also gratefully acknowledge the support of NVIDIA Corporation with the donation of a RTX A6000 GPU with 48Gb. The authors also thankfully acknowledge the grant of the Universidad de M\'alaga and the Instituto de Investigaci\'on Biom\'edica de M\'alaga y Plataforma en Nanomedicina-IBIMA Plataforma BIONAND.

\printbibliography

@article{ovalle2022hybrid,
  title={Hybrid classical--quantum Convolutional Neural Network for stenosis detection in X-ray coronary angiography},
  author={Ovalle-Magallanes, Emmanuel and Avina-Cervantes, Juan Gabriel and Cruz-Aceves, Ivan and Ruiz-Pinales, Jose},
  journal={Expert Systems with Applications},
  volume={189},
  pages={116112},
  year={2022},
  publisher={Elsevier}
}

@article{grandini2020metrics,
  title={Metrics for multi-class classification: an overview},
  author={Grandini, Margherita and Bagli, Enrico and Visani, Giorgio},
  journal={arXiv preprint arXiv:2008.05756},
  year={2020}
}

@article{wang2021review,
  title={A review of deep learning on medical image analysis},
  author={Wang, Jian and Zhu, Hengde and Wang, Shui-Hua and Zhang, Yu-Dong},
  journal={Mobile Networks and Applications},
  volume={26},
  pages={351--380},
  year={2021},
  publisher={Springer}
}

@article{rawat2017deep,
  title={Deep convolutional neural networks for image classification: A comprehensive review},
  author={Rawat, Waseem and Wang, Zenghui},
  journal={Neural computation},
  volume={29},
  number={9},
  pages={2352--2449},
  year={2017},
  publisher={MIT Press}
}

@inproceedings{huang2017densely,
  title={Densely connected convolutional networks},
  author={Huang, Gao and Liu, Zhuang and Van Der Maaten, Laurens and Weinberger, Kilian Q},
  booktitle={Proceedings of the IEEE conference on computer vision and pattern recognition},
  pages={4700--4708},
  year={2017}
}

@inproceedings{sandler2018mobilenetv2,
  title={Mobilenetv2: Inverted residuals and linear bottlenecks},
  author={Sandler, Mark and Howard, Andrew and Zhu, Menglong and Zhmoginov, Andrey and Chen, Liang-Chieh},
  booktitle={Proceedings of the IEEE conference on computer vision and pattern recognition},
  pages={4510--4520},
  year={2018}
}

@inproceedings{sae2019convolutional,
  title={Convolutional neural networks using MobileNet for skin lesion classification},
  author={Sae-Lim, Wannipa and Wettayaprasit, Wiphada and Aiyarak, Pattara},
  booktitle={2019 16th international joint conference on computer science and software engineering (JCSSE)},
  pages={242--247},
  year={2019},
  organization={IEEE}
}

@inproceedings{reddy2018comparison,
  title={Comparison of deep learning models for biometric-based mobile user authentication},
  author={Reddy, Narsi and Rattani, Ajita and Derakhshani, Reza},
  booktitle={2018 IEEE 9th international conference on biometrics theory, applications and systems (BTAS)},
  pages={1--6},
  year={2018},
  organization={IEEE}
}

@inproceedings{he2016deep,
  title={Deep residual learning for image recognition},
  author={He, Kaiming and Zhang, Xiangyu and Ren, Shaoqing and Sun, Jian},
  booktitle={Proceedings of the IEEE conference on computer vision and pattern recognition},
  pages={770--778},
  year={2016}
}

@article{rigatelli2022modern,
  title={Modern atlas of invasive coronary angiography views: a practical approach for fellows and young interventionalists},
  author={Rigatelli, Gianluca and Gianese, Filippo and Zuin, Marco},
  journal={The International Journal of Cardiovascular Imaging},
  volume={38},
  number={5},
  pages={919--926},
  year={2022},
  publisher={Springer}
}

@inproceedings{zhou2021review,
  title={Review of Vessel Segmentation and Stenosis classification in X-ray Coronary Angiography},
  author={Zhou, Ying and Guo, Haopan and Song, Jiarui and Chen, Yan and Wang, Jinjia},
  booktitle={2021 13th International Conference on Wireless Communications and Signal Processing (WCSP)},
  pages={1--5},
  year={2021},
  organization={IEEE}
}

@article{knuuti_2019_2020,
    author = {Knuuti, Juhani and Wijns, William and Saraste, Antti and Capodanno, Davide and Barbato, Emanuele and Funck-Brentano, Christian and Prescott, Eva and Storey, Robert F and Deaton, Christi and Cuisset, Thomas and Agewall, Stefan and Dickstein, Kenneth and Edvardsen, Thor and Escaned, Javier and Gersh, Bernard J and Svitil, Pavel and Gilard, Martine and Hasdai, David and Hatala, Robert and Mahfoud, Felix and Masip, Josep and Muneretto, Claudio and Valgimigli, Marco and Achenbach, Stephan and Bax, Jeroen J and ESC Scientific Document Group },
    title = "{2019 ESC Guidelines for the diagnosis and management of chronic coronary syndromes: The Task Force for the diagnosis and management of chronic coronary syndromes of the European Society of Cardiology (ESC)}",
    journal = {European Heart Journal},
    volume = {41},
    number = {3},
    pages = {407-477},
    year = {2019},
    month = {08},
    issn = {0195-668X},
    doi = {10.1093/eurheartj/ehz425},
    url = {https://doi.org/10.1093/eurheartj/ehz425},
    eprint = {https://academic.oup.com/eurheartj/article-pdf/41/3/407/32651471/ehz425.pdf},
}

@article{collet_2020_2021,
	title = {2020 {ESC} {Guidelines} for the management of acute coronary syndromes in patients presenting without persistent {ST}-segment elevation},
	volume = {42},
	issn = {0195-668X, 1522-9645},
	url = {https://academic.oup.com/eurheartj/article/42/14/1289/5898842},
	doi = {10.1093/eurheartj/ehaa575},
	language = {en},
	number = {14},
	urldate = {2021-07-30},
	journal = {European Heart Journal},
	author = {Collet, Jean-Philippe and Thiele, Holger and Barbato, Emanuele and Barthélémy, Olivier and Bauersachs, Johann and Bhatt, Deepak L and Dendale, Paul and Dorobantu, Maria and Edvardsen, Thor and Folliguet, Thierry and Gale, Chris P and Gilard, Martine and Jobs, Alexander and Jüni, Peter and Lambrinou, Ekaterini and Lewis, Basil S and Mehilli, Julinda and Meliga, Emanuele and Merkely, Béla and Mueller, Christian and Roffi, Marco and Rutten, Frans H and Sibbing, Dirk and Siontis, George C M and Chettibi, Mohammed and Hayrapetyan, Hamlet G and Metzler, Bernhard and Najafov, Ruslan and Stelmashok, Valeriy I and Claeys, Marc and Kušljugić, Zumreta and Gatzov, Plamen Marinov and Skoric, Bosko and Panayi, Georgios and Mates, Martin and Sorensen, Rikke and Shokry, Khaled and Marandi, Toomas and Kajander, Olli A and Commeau, Philippe and Aladashvili, Alexander and Massberg, Steffen and Nikas, Dimitrios and Becker, Dávid and Gudmundsd\'ottir, Ingibjörg J and Peace, Aaron J and Beigel, Roy and Indolfi, Ciro and Aidargaliyeva, Nazipa and Elezi, Shpend and Beishenkulov, Medet and Maca, Aija and Gustiene, Olivija and Degrell, Philippe and Cassar Maempel, Andrew and Ivanov, Victoria and Damman, Peter and Kedev, Sasko and Steigen, Terje K and Legutko, Jacek and Morais, João and Vinereanu, Dragos and Duplyakov, Dmitry and Zavatta, Marco and Pavlović, Milan and Orban, Marek and Bunc, Matjaž and Ibañez, Borja and Hofmann, Robin and Gaemperli, Oliver and Marjeh, Yassin Bani and Addad, Faouzi and Tutar, Eralp and Parkhomenko, Alexander and Karia, Nina and ESC Scientific Document Group},
	month = apr,
	year = {2021},
	pages = {1289--1367}
}

@article{zir1976interobserver,
  title={Interobserver variability in coronary angiography.},
  author={Zir, LEONARD M and Miller, STEPHEN W and Dinsmore, ROBERT E and Gilbert, JP and Harthorne, JW},
  journal={Circulation},
  volume={53},
  number={4},
  pages={627--632},
  year={1976},
  publisher={Am Heart Assoc}
}

@article{leape2000effect,
  title={Effect of variability in the interpretation of coronary angiograms on the appropriateness of use of coronary revascularization procedures},
  author={Leape, Lucian L and Park, Rolla Edward and Bashore, Thomas M and Harrison, J Kevin and Davidson, Charles J and Brook, Robert H},
  journal={American Heart Journal},
  volume={139},
  number={1},
  pages={106--113},
  year={2000},
  publisher={Elsevier}
}

@article{ovalle2022improving,
  title={Improving convolutional neural network learning based on a hierarchical bezier generative model for stenosis detection in X-ray images},
  author={Ovalle-Magallanes, Emmanuel and Avina-Cervantes, Juan Gabriel and Cruz-Aceves, Ivan and Ruiz-Pinales, Jose},
  journal={Computer Methods and Programs in Biomedicine},
  volume={219},
  pages={106767},
  year={2022},
  publisher={Elsevier}
}

@article{pang2021stenosis,
  title={Stenosis-DetNet: Sequence consistency-based stenosis detection for X-ray coronary angiography},
  author={Pang, Kun and Ai, Danni and Fang, Huihui and Fan, Jingfan and Song, Hong and Yang, Jian},
  journal={Computerized Medical Imaging and Graphics},
  volume={89},
  pages={101900},
  year={2021},
  publisher={Elsevier}
}

@article{danilov2021real,
  title={Real-time coronary artery stenosis detection based on modern neural networks},
  author={Danilov, Viacheslav V and Klyshnikov, Kirill Yu and Gerget, Olga M and Kutikhin, Anton G and Ganyukov, Vladimir I and Frangi, Alejandro F and Ovcharenko, Evgeny A},
  journal={Scientific reports},
  volume={11},
  number={1},
  pages={1--13},
  year={2021},
  publisher={Springer}
}

@article{moon2021automatic,
  title={Automatic stenosis recognition from coronary angiography using convolutional neural networks},
  author={Moon, Jong Hak and Cha, Won Chul and Chung, Myung Jin and Lee, Kyu-Sung and Cho, Baek Hwan and Choi, Jin Ho and others},
  journal={Computer methods and programs in biomedicine},
  volume={198},
  pages={105819},
  year={2021},
  publisher={Elsevier}
}

@article{zhou2021automated,
  title={Automated deep learning analysis of angiography video sequences for coronary artery disease},
  author={Zhou, Chengyang and Dinh, Thao Vy and Kong, Heyi and Yap, Jonathan and Yeo, Khung Keong and Lee, Hwee Kuan and Liang, Kaicheng},
  journal={arXiv preprint arXiv:2101.12505},
  year={2021}
}

@article{litjens2019state,
  title={State-of-the-art deep learning in cardiovascular image analysis},
  author={Litjens, Geert and Ciompi, Francesco and Wolterink, Jelmer M and de Vos, Bob D and Leiner, Tim and Teuwen, Jonas and I{\v{s}}gum, Ivana},
  journal={JACC: Cardiovascular imaging},
  volume={12},
  number={8 Part 1},
  pages={1549--1565},
  year={2019},
  publisher={American College of Cardiology Foundation Washington, DC}
}

@article{cai2020review,
  title={A review of the application of deep learning in medical image classification and segmentation},
  author={Cai, Lei and Gao, Jingyang and Zhao, Di},
  journal={Annals of translational medicine},
  volume={8},
  number={11},
  year={2020},
  publisher={AME Publications}
}

@article{zhou2021reviewDL,
  title={A review of deep learning in medical imaging: Imaging traits, technology trends, case studies with progress highlights, and future promises},
  author={Zhou, S Kevin and Greenspan, Hayit and Davatzikos, Christos and Duncan, James S and Van Ginneken, Bram and Madabhushi, Anant and Prince, Jerry L and Rueckert, Daniel and Summers, Ronald M},
  journal={Proceedings of the IEEE},
  volume={109},
  number={5},
  pages={820--838},
  year={2021},
  publisher={IEEE}
}

\newpage
\section*{Supplementary material} \label{sec:supplementary}
As described in the main manuscript, the test sets of the proposed experiments have different sizes since positive classes are cumulative,  which increases the number of samples used. Therefore, an additional and more restrictive test was carried out, where all test sets were equalized \cmt{established problems use the same test size. To do it, test sets were randomly reduced} reducing randomly to the lowest number of patches of the test sets, in this case, 62 patches in the 100\% lesion category. The F-measure and Area Under Curve (AUC) obtained are provided in \tablename s\  \ref{tab:F-test31} and \ref{tab:AUC-test31}, respectively. \cmt{,  and discussed in Section 3 of the main manuscript.} The first remarkable fact is, despite the fairer tests to compare, the values attained for both measures, F-measure and AUC, trends and behaviors are similar to the original test reported in Section 3 of the main manuscript. Besides, the outcomes for F-measure, \tablename \  \ref{tab:F-test31}, are lower than those obtained for AUC, \tablename \  \ref{tab:AUC-test31}. This fact could be explained as the F-measure considers Precision and Recall, where positive class classification performance is estimated, whereas AUC considers Precision and Specificity, which involves the performance of both classes. \cmt{measure how well classified both classes are.} These higher values for AUC could be interpreted as a slightly better classification of “non-lesion” patches despite both classes being balanced. The results are a bit spread, with $\geq$99\% lesions, F-measure attains $>$93\% of performance, and AUC achieves $>$98\%, while considering $<$99\% lesions, around 70\% and 75\% are obtained for F-measure and AUC, resp. Besides this considerable decrease when [98\%, 90\%] lesions are incorporated, there is a gradual growth as moderate lesions are considered; the enlargement of training patches could explain it. Finally, the values experienced a slight decline that could be interpreted as the more complex distinguishing mild lesions.

\begin{table*}
    \centering
    \caption{F-measure obtained on the test set using 5-fold stratified cross-validation, for the four strategies and the same size for the test sets. The highest values by rows are shown in bold.}
    \resizebox{\textwidth}{!}{
    \begin{tabular}{|c|c|ccccc|}
    \hline

    \multirow{2}{*}{Strategy} &\multirow{2}{*}{\shortstack{Lesion \\ Range}} & \multicolumn{5}{c|}{Convolutional Neural Networks Models} \\ 
                                                       
    & & DenseNet-201 & MobileNet-V2 & NasNet-Mobile & ResNet-18 & ResNet-50 \\
    \hline
    \multirow{7}{*}{\shortstack{With \\ 100\% and \\ 99\% lesions}} 
    & 100\% & $0.897 \pm 0.019$ & $0.866 \pm 0.091$ & $0.884 \pm 0.056$ & $0.853 \pm 0.089$ & $\boldsymbol{0.920 \pm 0.031}$ \\
    & $\geq 99\%$ & $ 0.900\pm0.033 $ & $ 0.927\pm 0.038$ & $0.896 \pm0.070 $ & $ 0.906\pm 0.091$ & $\boldsymbol{0.934 \pm0.055} $ \\
    & $\geq 90\%$ & $0.599 \pm 0.058$ & $0.537 \pm 0.041$ & $\boldsymbol{0.629 \pm 0.049}$ & $0.602 \pm 0.057$ & $0.549 \pm 0.065$ \\
    & $\geq 70\%$ & $0.661\pm 0.110$ & $\boldsymbol{0.744 \pm 0.088}$ & $0.707 \pm 0.018$ & $0.698 \pm 0.056$ & $0.680 \pm 0.047$ \\
    & $\geq 50\%$ & $0.534 \pm 0.126$ & $\boldsymbol{0.630 \pm 0.072}$ & $0.527 \pm 0.102$ & $0.511 \pm 0.040$ & $0.516 \pm 0.094$ \\
    & $\geq 20\%$ & $0.706 \pm 0.031$ & $\boldsymbol{0.727 \pm 0.064}$ & $0.721 \pm 0.037$ & $0.678 \pm 0.077$ & $0.694 \pm 0.079$ \\
    & $>$ 0\% & $0.669 \pm 0.088$ & $\boldsymbol{0.705 \pm 0.092}$ & $0.694 \pm 0.062$ & $0.651 \pm 0.043$ & $0.669 \pm 0.043$ \\
    \hline
    \multirow{7}{*}{\shortstack{With \\ 100\% and \\ 99\% lesions \\+ Data \\ Augmentation}} 
    & 100\% & $0.894 \pm 0.034$ & $0.893 \pm 0.098$ & $0.803 \pm 0.081$ & $\boldsymbol{0.927 \pm 0.032}$ & $0.914 \pm 0.060$ \\
    & $\geq 99\%$ & $0.851 \pm 0.157$ & $0.799 \pm 0.253$ & $\boldsymbol{0.895 \pm 0.036}$ & $0.744 \pm 0.299$ & $0.752 \pm 0.225$\\
    & $\geq 90\%$ & $\boldsymbol{0.597 \pm 0.066}$ & $0.572 \pm 0.070$ & $0.564 \pm 0.038$ & $0.596 \pm 0.087$ & $0.529 \pm 0.072$ \\
    & $\geq 70\%$ & $0.699 \pm 0.024$ & $\boldsymbol{0.728 \pm 0.053}$ & $0.707 \pm 0.038$ & $0.680 \pm 0.059$ & $0.657 \pm 0.065$ \\
    & $\geq 50\%$ & $0.533 \pm 0.103$ & $\boldsymbol{0.648 \pm 0.056}$ & $0.537 \pm 0.071$ & $0.470 \pm 0.110$ & $0.506 \pm 0.138$ \\
    & $\geq 20\%$ & $0.671 \pm 0.025$ & $0.645 \pm 0.055$ & $\boldsymbol{0.691 \pm 0.056}$ & $0.673 \pm 0.042$ & $0.669 \pm 0.054$ \\
    & $>$ 0\% & $0.651 \pm 0.098$ & $0.650 \pm 0.069$ & $\boldsymbol{0.692 \pm 0.032}$ & $0.657 \pm 0.056$ & $0.569 \pm 0.066$ \\
    \hline
    \multirow{5}{*}{\shortstack{W/o \\ 100\% and \\ 99\% lesions}}
    & $\geq 90\%$ & $\boldsymbol{0.627 \pm 0.171}$ & $0.568 \pm 0.089$ & $0.603 \pm 0.130$ & $0.625 \pm 0.079$ & $0.490 \pm 0.093$ \\
    & $\geq 70\%$ & $\boldsymbol{0.781 \pm 0.068}$ & $0.621 \pm 0.101$ & $0.684 \pm 0.090$ & $0.682 \pm 0.097$ & $0.642 \pm 0.103$ \\
    & $\geq 50\%$ & $0.667 \pm 0.050$ & $\boldsymbol{0.697 \pm 0.090}$ & $0.663 \pm 0.053$ & $0.694 \pm 0.050$ & $0.689 \pm 0.116$ \\
    & $\geq 20\%$ & $0.634 \pm 0.055$ & $0.593 \pm 0.067$ & $0.627 \pm 0.069$ & $0.621 \pm 0.083$ & $\boldsymbol{0.642 \pm 0.087}$ \\
    & $>$ 0\% & $0.583 \pm 0.096$ & $\boldsymbol{0.633 \pm 0.057}$ & $0.631 \pm 0.085$ & $0.588 \pm 0.056$ & $0.560 \pm 0.055$ \\
    \hline
    \multirow{5}{*}{\shortstack{W/o \\ 100\% and \\ 99\% lesions \\+ Data \\ Augmentation}}
    & $\geq 90\%$ & $0.462 \pm 0.088$ & $\boldsymbol{0.552 \pm 0.136}$ & $0.529 \pm 0.119$ & $0.463 \pm 0.175$ & $0.289 \pm 0.091$ \\
    & $\geq 70\%$ & $0.650 \pm 0.109$ & $0.660 \pm 0.072$ & $\boldsymbol{0.671 \pm 0.050}$ & $0.612 \pm 0.084$ & $0514 \pm 0.043$ \\
    & $\geq 50\%$ & $\boldsymbol{0.769 \pm 0.072}$ & $0.755 \pm 0.056$ & $0.743 \pm 0.040$ & $0.704 \pm 0.033$ & $0.704 \pm 0.036$ \\
    & $\geq 20\%$ & $0.608 \pm 0.049$ & $0.639 \pm 0.076$ & $\boldsymbol{0.687 \pm 0.054}$ & $0.610 \pm 0.027$ & $0.637 \pm 0.096$ \\
    & $>$ 0\% & $0.593 \pm 0.042$ & $\boldsymbol{0.676 \pm 0.081}$ & $0.674 \pm 0.020$ & $0.601 \pm 0.097$ & $0.627 \pm 0.079$ \\
    \hline
    \end{tabular}
    } \label{tab:F-test31}
\end{table*}

\begin{table*}
    \centering
    \caption{AUC obtained on the test set using 5-fold stratified cross-validation, for the four strategies and the same size for the test sets. The highest values by rows are shown in bold.}
    \resizebox{\textwidth}{!}{
    \begin{tabular}{|c|c|ccccc|}
    \hline

    \multirow{2}{*}{Strategy} &\multirow{2}{*}{\shortstack{Lesion \\ Range}} & \multicolumn{5}{c|}{Convolutional Neural Networks Models} \\ 
                                                       
    & & DenseNet-201 & MobileNet-V2 & NasNet-Mobile & ResNet-18 & ResNet-50 \\
    \hline
    \multirow{7}{*}{\shortstack{With \\ 100\% and \\ 99\% lesions}} 
    & 100\% & $\boldsymbol{0.971 \pm 0.023}$ & $0.950 \pm 0.035$ & $0.943 \pm 0.039$ & $0.969 \pm 0.020$ & $0.958 \pm 0.013$ \\
    & $\geq 99\%$ & $ \boldsymbol{0.983\pm 0.011 }$ & $ 0.963\pm 0.030$ & $0.952 \pm 0.068 $ & $0.980 \pm 0.024 $ & $0.979 \pm 0.018 $ \\
    & $\geq 90\%$ & $0.738 \pm 0.044$ & $0.694 \pm 0.062$ & $0.736 \pm 0.037$ & $0.729 \pm 0.047$ & $\boldsymbol{0.745 \pm 0.036}$ \\
    & $\geq 70\%$ & $0.834 \pm 0.088$ & $\boldsymbol{0.863 \pm 0.048}$ & $0.846 \pm 0.032$ & $0.833 \pm 0.089$ & $0.813 \pm 0.052$ \\
    & $\geq 50\%$ & $0.796 \pm 0.058$ & $\boldsymbol{0.817 \pm 0.021}$ & $0.719 \pm 0.054$ & $0.756 \pm 0.034$ & $0.718 \pm 0.068$ \\
    & $\geq 20\%$ & $0.820 \pm 0.045$ & $0.795 \pm 0.048$ & $0.819 \pm 0.053$ & $0.807 \pm 0.020$ & $\boldsymbol{0.821 \pm 0.043}$ \\
    & $>$ 0\% & $0.805 \pm 0.052$ & $0.816 \pm 0.047$ &$\boldsymbol{0.819 \pm 0.023}$ & $0.774 \pm 0.032$ & $0.770 \pm 0.023$ \\
    \hline
    \multirow{7}{*}{\shortstack{With \\ 100\% and \\ 99\% lesions \\+ Data \\ Augmentation}}
    & 100\% & $0.949 \pm 0.022$ & $0.924 \pm 0.073$ & $0.914 \pm 0.052$ & $\boldsymbol{0.973 \pm 0.017}$ & $0.973 \pm 0.026$ \\
    & $\geq 99\%$ & $0.933 \pm 0.092$ & $0.861 \pm 0.222$ & $\boldsymbol{0.947 \pm 0.022}$ & $0.893 \pm 0.118$ & $0.869 \pm 0.144$ \\
    & $\geq 90\%$ & $0.709 \pm 0.034$ & $\boldsymbol{0.715 \pm 0.040}$ & $0.689 \pm 0.043$ & $0.698 \pm 0.049$ & $0.680 \pm 0.065$ \\
    & $\geq 70\%$ & $0.829 \pm 0.020$ & $\boldsymbol{0.834 \pm 0.025}$ & $0.831 \pm 0.058$ & $0.810 \pm 0.059$ & $0.817 \pm 0.073$ \\
    & $\geq 50\%$ & $\boldsymbol{0.796 \pm 0.033}$ & $0.749 \pm 0.011$ & $0.753 \pm 0.049$ & $0.690 \pm 0.073$ & $0.677 \pm 0.066$ \\
    & $\geq 20\%$ & $0.791 \pm 0.030$ & $0.753 \pm 0.029$ & $0.788 \pm 0.037$ & $\boldsymbol{0.820 \pm 0.015}$ & $0.760 \pm 0.030$ \\
    & $>$ 0\% & $0.777 \pm 0.056$ & $0.778 \pm 0.025$ & $\boldsymbol{0.811 \pm 0.035}$ & $0.788 \pm 0.033$ & $0.758 \pm 0.041$ \\
    \hline
    \multirow{5}{*}{\shortstack{W/o \\ 100\% and \\ 99\% lesions}}
    & $\geq 90\%$ & $0.811 \pm 0.034$ & $0.776 \pm 0.040$ & $0.764 \pm 0.066$ & $\boldsymbol{0.819 \pm 0.033}$ & $0.740 \pm 0.073$ \\
    & $\geq 70\%$ & $\boldsymbol{0.816 \pm 0.058}$ & $0.776 \pm 0.086$ & $0.794 \pm 0.015$ & $0.805 \pm 0.047$ & $0.791 \pm 0.037$ \\
    & $\geq 50\%$ & $0.843 \pm 0.035$ & $\boldsymbol{0.858 \pm 0.051}$ & $0.827 \pm 0.065$ & $0.833 \pm 0.028$ & $0.827 \pm 0.085$ \\
    & $\geq 20\%$ & $\boldsymbol{0.784 \pm 0.023}$ & $0.742 \pm 0.036$ & $0.752 \pm 0.017$ & $0.751 \pm 0.057$ & $0.765 \pm 0.049$ \\
    & $>$ 0\% & $\boldsymbol{0.778 \pm 0.059}$ & $0.758 \pm 0.035$ & $0.776 \pm 0.035$ & $0.757 \pm 0.042$ & $0.747 \pm 0.032$ \\
    \hline
    \multirow{5}{*}{\shortstack{W/o \\ 100\% and \\ 99\% lesions \\+ Data \\ Augmentation}}
    & $\geq 90\%$ & $0.713 \pm 0.071$ & $\boldsymbol{0.754 \pm 0.078}$ & $0.733 \pm 0.116$ & $0.702 \pm 0.114$ & $0.664 \pm 0.100$ \\
    & $\geq 70\%$ & $0.773 \pm 0.077$ & $0.749 \pm 0.027$ & $\boldsymbol{0.779 \pm 0.051}$ & $0.776 \pm 0.011$ & $0.713 \pm 0.037$ \\
    & $\geq 50\%$ & $\boldsymbol{0.863 \pm 0.023}$ & $0.856 \pm 0.030$ & $0.851 \pm 0.027$ & $0.829 \pm 0.047$ & $0.846 \pm 0.043$ \\
    & $\geq 20\%$ & $0.772 \pm 0.049$ & $0.752 \pm 0.035$ & $0.763 \pm 0.036$ & $\boldsymbol{0.776 \pm 0.023}$ & $0.743 \pm 0.049$ \\
    & $>$ 0\% & $0.767 \pm 0.036$ & $\boldsymbol{0.789 \pm 0.046}$ & $0.774 \pm 0.046$ & $0.743 \pm 0.088$ & $0.765 \pm 0.052$ \\
    \hline
    \end{tabular}
    } \label{tab:AUC-test31}
\end{table*}

\vspace{3em}

\end{document}